# Towards a Comprehensive Numerical Scheme for Simulating Incompressible Two-Phase Flows

Jun-De Li
College of Engineering and Science,
Victoria University
Australia
Tel: +61-3-99194105
Email: *Jun-De.Li@vu.edu.au*

**Abstract** A comprehensive scheme for the spatial discretisation of continuity equation, momentum advection and normal and shear stresses at the fluid interfaces is presented for numerically simulating the incompressible two phase flows based on the finite volume and volume of fluid methods. The scheme uses the mass conservation for the advection of the interface between the two fluids, physical considerations in determining the momentum flux at the surfaces of the finite volume so that it is consistent with that for the advection of the interfaces, and correctly accounts for the contributions to the normal and shear stresses from the viscosities of the two fluids.

An interface reconstruction method is also presented, which approximates the local interface by a second order surface. The reconstruction method is based on the local average height function. It is linear and mathematically consistent. The force due to the surface tension is calculated based on the local surface area, curvature and unit normal of the interface which can all be calculated analytically from the reconstructed interface.

## 1. Introduction

In many natural phenomenon and engineering applications, fluids with large differences in density and viscosity co-exist in a flow field, such as bubbles rising in boiling processes, atomization of water jets into small droplets from sprinkler systems used in fire suppressions, waves on the sea, waterfalls, etc. In these flows, not only the fluids have large differences in their physical properties, but also complex evolving interfaces exist between the fluids. At these interfaces, surface tension can play a critical role in controlling the local behaviour of the fluid motions.

In numerical simulations of single phase fluid flows, the Navier-Stokes equations are generally solved numerically using different spatial discretisation schemes such as the finite volume scheme [1] or the finite element scheme [2]. In this work, we consider only the finite volume scheme. In the simulations of two phase flows, on top of numerically solving the Navier-Stokes equations, the evolving complex interfaces need to be followed. This has so far been achieved by either the interface tracking [3-7] methods which follow the fluid particles on the interface and are Lagrangian in nature, or the interface capturing methods [8-14] which reconstruct the interface in the flow field at each time step and are Eulerian in nature. It has been found that the interface tracking methods work well mainly for multi-phase flows without significant topological changes of the domains occupied by each fluid. In many multi-phase flows, complex topological changes of the fluid domains occur regularly.

There are currently several interface capturing methods which have been used to reconstruct the interfaces during the numerical simulations of the two phase flows. These include the Volume of Fluid (VOF) method [14-16], Level Set (LS) method [17-25], Moment of Fluid (MOF) method [26-32], Coupled LS and VOF (CLSVOF) method [33, 34], Coupled LS and MOF (CLSMOF) method [35], and Coupled LS/VOF/ghost fluid method [36]. In these methods, the interface separating the two immiscible fluids is approximated as a liner function in each grid cell by generally using a Piecewise Linear Interface Calculation (PLIC) [36, 37]. Thus, these methods all approximate the interface locally as linear plane surface. As it has been pointed out in [38], "The interface is more and more poorly resolved as the simulation goes on. However, this is a major issue with every VOF methods that limits the interface reconstruction to a single linear equation in a cut cell".

In addition to capturing the interface, a model is also needed to calculate the forces due to the surface tension at the interfaces. In nearly all the interface capture methods, this is accounted for by a source term in the momentum equations and is generally calculated using the Continuous Surface Forces (CSF) model [39]. The CSF model depends on the approximation of the interface curvature from the gradient of the VOF (or LS) function. As pointed out in [40], the gradient cannot be calculated accurately in numerical simulations since VOF function is a discontinuous step function and its discrete approximations are known to generate unphysical spurious currents at the interface. Physically, forces due to surface tension exist only on curved surfaces. Renardy and Renardy [41] have developed the Parabolic Reconstruction Of Surface Tension (PROST) method which was shown to achieve a significant reduction in the magnitude of the velocity induced by the spurious currents. The model involves a nonlinear least square optimization and many iterations are needed in each construction of the interface. It was noted that some exceptions occur as breakup is reached and in these cases, the optimization process was interrupted after 200 iteration steps. In this work, we will present a Local Second Order Curve (LSOC) fitting to the interface through a least square optimization by solving a set of linear equations (even through the fitted surface is second order) and calculate the forces due to the surface tension over the locally curved surface. There will be no iterations required in the reconstruction of the interface and the curvature can be determined analytically. The surface fitting scheme is similar to the parabola-fitted surface scheme given in [42], but it is different from it in that a local integration is preformed to ensure that the average height function is approximated in a column. Without performing this local integration, the height function is implicitly assumed to be positioned at the centre of the column, but in reality interfaces may not even cut through the centre of the column.

The reconstruction of the interface is in general accompanied by a numerical scheme for the advection of the fluid interface [26, 33, 38, 42]. The objectives of this advection scheme are to produce a sharp interface between the two fluids and to ensure mass conservation. To meet these objectives, the mass advections in each of the faces of a finite volume need to be carefully determined. However, it seems that this advection scheme developed for the VOF interface has not been applied to the momentum equations in general. Since any mass flows carry with it the momentum, it is expected that the advection scheme developed for the interface should also be applied accordingly to that of the momentum equations where an interface exists. Otherwise, the scheme for the advection of the interface and that for the advection of momentum will not be consistent. We will present methods of calculating the advected masses on the surfaces of the finite volume and these masses will be used in calculating the momentum advections, thus ensuring the discretisation of the mass conservation and that of the momentum equations are consistent.

By applying the finite volume scheme, the forces due to viscosity generating the normal and shear stresses on the cell surfaces which intersect with the interface need to be calculated. In this case, the stress calculations on these cell surfaces cannot be simply that of using a viscosity averaged based on volume fraction in order to take into account the viscosity jump appropriately at the interface. It is found that the surface areas intercepted by the fluids of different phases are required. We will present the methods of calculating these surface areas from on the intersections of the interface with the cell surfaces.

In this work, we will denote the two fluids with large density and viscosity differences as the heavy and light fluids. The Navier-Stokes equations can be applied to both the heavy fluid and the light fluid, separately. In doing so, boundary conditions are required for both the heavy and light fluids at the interface (of course, these boundary conditions are connected between the heavy and light fluids). However, in simulating two phase flows, it has been a common practice to apply a one fluid model [43] which considers an equivalent fluid whose physical properties are established with the average properties of the flow through a volume of fluid $c$,

$$\begin{aligned} \rho(c) &= \rho_H c + (1-c)\rho_L \\ \mu(c) &= \mu_H c + (1-c)\mu_L \end{aligned} \tag{1}$$

where $\rho_H$ ($\mu_H$) is the density (viscosity) of the heavy fluid, $\rho_L$ ($\mu_L$) is the density (viscosity) of the light fluid and $\rho$ ($\mu$) is the equivalent (or volume average) density (viscosity). The volume of fluid $c = 1$ for the heavy fluid, $c = 0$ for the light fluid and $0 < c < 1$ in regions where the two fluids meet, i.e. where the interface is located. In (1), the equivalent viscosity is based on the volume weighted average and we will show that this may not be an appropriate weighting for two fluids with well-defined interfaces between them.

The Navier-Stokes equations for one fluid with interfaces between the heavy fluid and light fluid have generally been expressed as:

$$\begin{aligned} \frac{\partial \rho \mathbf{u}}{\partial t} + \nabla .(\rho \mathbf{u} \otimes \mathbf{u}) &= -\nabla p + \nabla .(2\mu \mathbf{D}) + \sigma \kappa \delta_s \mathbf{n} + \mathbf{g}(\rho - \rho_{ref}) \\ \frac{\partial \rho}{\partial t} + \nabla .(\rho \mathbf{u}) &= 0 \\ \nabla .\mathbf{u} &= 0 \end{aligned} \tag{2}$$

Here $\mathbf{u}$ is the velocity vector, $p$ is the pressure, $\mu$ is the viscosity, $D = (\nabla \mathbf{u} + \nabla \mathbf{u}^T)/2$ is the strain rate tensor, $\sigma$ is the surface tension, $\kappa$ is the mean curvature, $\mathbf{n}$ is the interface unit normal, $\delta_s$ is a delta function which identifies the interface location, $\mathbf{g}$ is the gravity acceleration and $\rho_{ref}$ is the reference density. The last term in the above momentum equation is the buoyance force which depends on the choice of the reference density $\rho_{ref}$. For example, for bubbles rising in a boiling process, $\rho_{ref} = \rho_H$ while for the atomization of water jets into small droplets from sprinkler systems used in fire suppressions, it is common to take $\rho_{ref} = \rho_L$.

The volume of fluid satisfies

$$\frac{\partial c}{\partial t} + \nabla.(c\mathbf{u}) = 0 \qquad (3)$$

This equation is the same as that for the density and thus only one of these equations needs to be solved in practice.

## 2. Time discretization

Here we consider only unsteady flows. The discretisation of the volume of fluid/density and pressure in time is similar to that in [42] and leads to

$$\frac{1}{\Delta t}[(\rho\mathbf{u})^{n+1} - (\rho\mathbf{u})^n] + [\nabla.(\rho\mathbf{u}\otimes\mathbf{u})]^{n+1/2} = -\nabla p^{n+1/2} \\ + \nabla[\mu^{n+1/2}(\mathbf{D}^n + \mathbf{D}^{n+1})] + (\sigma\kappa\delta_s\mathbf{n})^{n+1/2} + \mathbf{g}(\rho^{n+1/2} - \rho_{ref}) \qquad (4)$$

$$\frac{\rho^{n+1} - \rho^n}{\Delta t} + \nabla.(\rho\mathbf{u})^{n+1/2} = 0 \qquad (5)$$

$$\nabla.\mathbf{u}^{n+1} = 0 \qquad (6)$$

Because of the nonlinearity of the Navier-Stokes equations, the above equations (4-6) can in general be solved iteratively at each time step. An iteration loop in finding the solution at each time step consists of the following steps (1)–(4):

(1) Calculate the intermediate density $\rho^*$

$$\frac{\rho^* - \rho^n}{\Delta t} + \nabla.(\rho\mathbf{u})^{n+1/2} = 0 \qquad (7)$$

(2) Calculate the intermediate velocity $\mathbf{u}^*$

$$\mathbf{u}^* - \frac{\Delta t}{\rho^*}\nabla[\mu^{n+1/2}\mathbf{D}^*] = \frac{1}{\rho^*}(\rho\mathbf{u})^n + \frac{\Delta t}{\rho^*}\{-[\nabla.(\rho\mathbf{u}\otimes\mathbf{u})]^{n+1/2} \\ + \nabla[\mu^{n+1/2}\mathbf{D}^n] + (\sigma\kappa\delta_s\mathbf{n})^{n+1/2} + \mathbf{g}(\rho^{n+1/2} - \rho_{ref})\} \qquad (8)$$

(3) Calculate the pressure

$$\nabla.(\frac{\Delta t}{\rho^*}\nabla p^{n+1/2}) = \nabla.\mathbf{u}^* \qquad (9)$$

(4) Calculate the velocity correction

$$\mathbf{u}^{n+1} = \mathbf{u}* - \frac{\Delta t}{\rho *}\nabla p^{n+1/2} \tag{10}$$

(5) Repeat the steps (1) to (4) if the solutions are not converged.

To start the iteration, we need $\rho^{n+1/2}$ and $\mathbf{u}^{n+1/2}$ which are not known a prior. If we take $\rho*$ to be an approximation of $\rho^{n+1}$, in steps (1) and (2) in the first iteration loop, the iteration starts with $\rho^{n+1/2} = \rho^n$ and then $\rho^{n+1/2} = 0.5(\rho^n + \rho*)$ in sequential iteration loops. Similarly, for $\mathbf{u}^{n+1/2}$, we can start with $\mathbf{u}^{n+1/2} = \mathbf{u}^n$ and $\mathbf{u}^{n+1/2} = 0.5(\mathbf{u}^{n+1} + \mathbf{u}^n)$ in sequential iteration loops. Finally, after the solutions have converged, we can set $\rho^{n+1} = \rho*$.

## 3. Finite volume spatial discretisation at the interface

Here we consider a Cartesian coordinate system (*x, y, z*) with the corresponding velocity components (*u, v, w*). We will use cells to represent small finite volumes (or control volumes). The cell index at the cell centre will be expressed using subscript $(i, j, k)$ and the surfaces between cells will be ½ of its index increment, e.g. (*i, j+1/2, k*) for the surface index between $(i, j, k)$ and $(i, j+1, k)$. Also, only the subscript index that changes will in general be shown, e.g. *i-1/2* will mean (*i-1/2, j, k*), etc. All the variables will be stored at the cell centres and those at the surfaces between cells can be worked out from interpolations. The discretisation of the buoyance force term in (4) is trivial and will not be shown here.

(1) The continuity equation

This can also be considered as the interface advection because of the close connection between the volume of fluid and the density (1). We use the continuity equation (mass conservation) because of its obvious physical meaning. By considering a finite volume over the cell (*i, j, k*) (Figure 1 shows only that varying in *i* and *j* indexes), the intermediate density $\rho*$ can be calculated as

$$\rho* = \rho^n + \frac{\Delta t}{V}[(\rho u)_{i-1/2}^{n+1/2} A_{i-1/2} - (\rho u)_{i+1/2}^{n+1/2} A_{i+1/2}$$
$$+ (\rho v)_{j-1/2}^{n+1/2} A_{j-1/2} - (\rho v)_{j+1/2}^{n+1/2} A_{j+1/2} + (\rho w)_{k-1/2}^{n+1/2} A_{k-1/2}$$
$$- (\rho w)_{k+1/2}^{n+1/2} A_{k+1/2}] \tag{11}$$
$$= \rho^n + \frac{1}{V}[(m_L + m_H)_{i-1/2}^{n+1/2} - (m_L + m_H)_{i+1/2}^{n+1/2} +$$
$$(m_L + m_H)_{j-1/2}^{n+1/2} - (m_L + m_H)_{j+1/2}^{n+1/2} + (m_L + m_H)_{k-1/2}^{n+1/2} - (m_L + m_H)_{k+1/2}^{n+1/2}]$$

Here $(m_L + m_H)_{i-1/2}^{n+1/2} = \Delta t(\rho u A)_{i-1/2}^{n+1/2}$ are the masses from the light and heavy fluids that have been advected from the upstream cell $(i-1, j, k)$ into the cell $(i, j, k)$ passing the surface $A_{i-1/2}$ during the time step $\Delta t$ as that shown in Figure 1, and V is the volume of the cell $(i, j, k)$. In the above discretisation, we have also assumed that the $u_{i-1/2}^{n+1/2}$ and $u_{i+1/2}^{n+1/2}$ are positive so that the scheme is upwind (similarly, $v_{j-1/2}^{n+1/2}$, $v_{j+1/2}^{n+1/2}$, $w_{k-1/2}^{n+1/2}$ and $w_{k+1/2}^{n+1/2}$ are assumed to be positive).

In case that $u_{i-1/2}^{n+1/2}$ and $u_{i+1/2}^{n+1/2}$ are negative, the terms for the advection of masses, e.g. in the x-direction, should be changed to $(m_L + m_H)_{i+1/2}^{n+1/2} - (m_L + m_H)_{i-1/2}^{n+1/2}$. Here $(m_L + m_H)_{i-1/2}^{n+1/2}$ is the mass leaving the cell $(i, j, k)$ from the surface $A_{i-1/2}$ and $(m_L + m_H)_{i+1/2}^{n+1/2}$ is the mass entering the cell $(i, j, k)$ through the surface $A_{i+1/2}$.

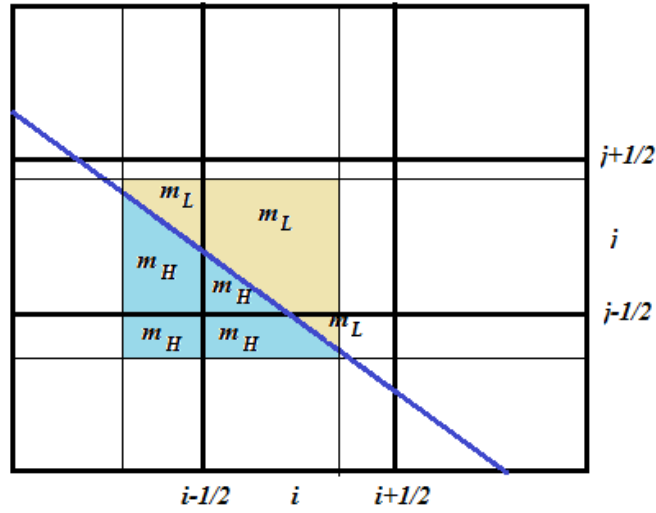

Figure 1 2D schematics showing the masses of the heavy and light fluids to flow into the centre cell $(i, j, k)$ during the time interval $\Delta t$.

From this calculation of the density at the cell $(i, j, k)$, the intermediate volume of fluid $c^*$ occupied by the heavy fluid can be calculated because $c^n$ is known and the volumes from $m_{H,i-1/2}$, $m_{H,i+1/2}$, $m_{H,j-1/2}$, $m_{H,j+1/2}$, $m_{H,k-1/2}$ and $m_{H,k+1/2}$ can all be calculated. This will allow the reconstruction of the interface (see below) at each iteration step.

*Remark 1:* In the literature, the advection of the interface has been accomplished using either directional split schemes [34, 38, 43] or unsplit schemes [45-50]. In a directional split scheme, the advection in the *x*, *y* and *z* directions are calculated sequentially. Figure 1 shows that, when using a directional split scheme, the interface needs to be reconstructed after the advection in each direction. Otherwise, mass conservation cannot be satisfied. There will be three interface reconstructions in 3D flows for each iteration loop. The unsplit scheme requires working out the amount of mass being advected into the cell $(i, j, k)$ from all its surrounding cells in a very complex manner [50] and only one reconstruction is required at each iteration loop. The directional split scheme has been commonly used because of its simplicity.

(2) The surface tension term is discretised as

$$\frac{\Delta t}{V}(\int_{A_s}\sigma\kappa\mathbf{n}dS)_{i,j,k} \qquad (12)$$

Here $A_s$ is the interface area inside the cell $(i,j,k)$ between the heavy and light fluids and $\mathbf{n}$ is the unit normal vector of the interface. $A_s=0$ if there is no interface inside the cell $(i,j,k)$. It should be noted that the force due to the surface tension is a force at the interface and its magnitude depends on the interface area $A_s$, curvature $\kappa$ and unit normal $\mathbf{n}$. Detail methods of calculating this force due to the surface tension will be presented later.

(3) The advection of momentum

Here we consider the *x*-direction discretisation only and those for the *y* and *z* directions are similar. By taking a volume integration of the momentum equation over the cell $(i,j,k)$ and using the Gauss Theorem to change the volume integration into the integrations over the cell surfaces, after multiplying every term in the momentum equation by $\Delta t/V$, the momentum enters and leaves the cell $(i,j,k)$ can be written as

$$\frac{\Delta t}{V}[(\rho uu)_{i-1/2}^{n+1/2}A_{i-1/2} - (\rho uu)_{i+1/2}^{n+1/2}A_{i+1/2} + (\rho uv)_{j-1/2}^{n+1/2}A_{j-1/2} \\ -(\rho uv)_{j+1/2}^{n+1/2}A_{j+1/2} + (\rho uw)_{k-1/2}^{n+1/2}A_{k-1/2} - (\rho uw)_{k+1/2}^{n+1/2}A_{k+1/2}] \qquad (13)$$

Assuming that $u_{i+1/2}^{n+1/2}$, $u_{i-1/2}^{n+1/2}$, $v_{j+1/2}^{n+1/2}$, $v_{j-1/2}^{n+1/2}$, $w_{k+1/2}^{n+1/2}$ and $w_{k-1/2}^{n+1/2}$ are all positive as before for an upwind scheme, the discretisation (13) can be written as

$$\frac{1}{V}[(m_L + m_H)_{i-1/2}^{n+1/2}u_{i-1/2}^{n+1/2} - (m_L + m_H)_{i+1/2}^{n+1/2}u_{i+1/2}^{n+1/2} + (m_L + m_H)_{j-1/2}^{n+1/2}u_{j-1/2}^{n+1/2} \\ -(m_L + m_H)_{j+1/2}^{n+1/2}u_{j+1/2}^{n+1/2} + (m_L + m_H)_{k-1/2}^{n+1/2}u_{k-1/2}^{n+1/2} - (m_L + m_H)_{k+1/2}^{n+1/2}u_{k+1/2}^{n+1/2}] \qquad (14)$$

Here, $(m_L + m_H)_{i-1/2}^{n+1/2}$ have the same meaning as in (11) and that shown in Figure 1, to be consistent with the advection of the interface as explained early, $u_{i-1/2}^{n+1/2}$ has been normally approximated as the interpolated surface centre velocity (MAC) between $u_{i-1,j,k}^{n+1/2}$ and $u_{i,j,k}^{n+1/2}$, $u_{j+1/2}^{n+1/2}$ is the interpolated velocity between $u_{i,j,k}^{n+1/2}$ and $u_{i,j+1,k}^{n+1/2}$, and $u_{k+1/2}^{n+1/2}$ is the interpolated velocity between $u_{i,j,k}^{n+1/2}$ and $u_{i,j,k+1}^{n+1/2}$. The physical meaning of each of the terms in (14) is very clear. For example, $(m_L + m_H)_{i-1/2}^{n+1/2}u_{i-1/2}^{n+1/2}$ is the momentum entering the cell $(i,j,k)$ from the upstream surface $A_{i-1/2}$ during the time period $\Delta t$.

*Remark 2*: $u_{i-1/2}^{n+1/2}$, $u_{j-1/2}^{n+1/2}$, $u_{k-1/2}^{n+1/2}$, $u_{i+1/2}^{n+1/2}$, $u_{j+1/2}^{n+1/2}$ and $u_{k+1/2}^{n+1/2}$ can be better approximated by the average velocities at the upstream volume centres occupied by the heavy and light fluids, e.g. $u_{i-1/2}^{n+1/2}$ can be better approximated by $u(x_{i-1/2} - 0.5*u_{i-1/2}\Delta t)$. This will use the velocity at approximately the volume centre location of $(m_L + m_H)_{i-1/2}^{n+1/2}$ as that shown in Figure 1 assuming $u_{i-1/2}^{n+1/2} > 0$. An even more accurate approximation to this velocity will be using the velocity at the mass centres $x_L$ and $x_H$ of $(m_L)_{i-1/2}^{n+1/2}$ and $(m_H)_{i-1/2}^{n+1/2}$ in cell $(i-1,j,k)$. The details of determining these quantities will be explained later.

*Remark 3*: In the above discretization for the momentum advection, the momentum entering the cell $(i,j,k)$, say, $(m_L + m_H)_{i-1/2}^{n+1/2} u_{i-1/2}^{n+1/2}$ is different from that calculated using the volume of fluid at the cell $(i-1, j, k)$ or that from an interpolation between those in cells $(i-1, j, k)$ and $(i,j,k)$. This is because the masses of the heavy and light fluids entering the cell $(i,j,k)$ during the time interval $\Delta t$ from the cell $(i-1, j, k)$ as that shown in Figure 1 will in general be different from that calculated using (1) because of the complexity of the interface orientation in general.

*Remark 4*: The discretization of the continuity equation (11) and the momentum advection terms (14) has automatically taken into account the density jump across the interface. This jump is finite due to the use of the finite volume method. On the other hand, direct differentiation will result in a delta function which is locally infinite and will result in larger errors during discretisation. This is one of the advantages of using finite volume method for the spatial discretisation.

(4) The viscous term $\nabla[\mu^{n+1/2}\mathbf{D}^n]$

Again, we consider only the discretisation in the *x*-direction and the other directions can be similarly performed. Using the same procedure as that in deriving (13), we have for the forces from the normal and shear stresses on the six surfaces of the cell $(i,j,k)$

$$\frac{\Delta t}{V}[(\mu A)_{i+1/2}^{n+1/2}(\frac{\partial u}{\partial x})_{i+1/2}^n - (\mu A)_{i-1/2}^{n+1/2}(\frac{\partial u}{\partial x})_{i-1/2}^n + (\mu A)_{j+1/2}^{n+1/2}(\frac{\partial u}{\partial y})_{j+1/2}^n$$
$$- (\mu A)_{j-1/2}^{n+1/2}(\frac{\partial u}{\partial y})_{j-1/2}^n + (\mu A)_{k+1/2}^{n+1/2}(\frac{\partial u}{\partial z})_{k+1/2}^n - (\mu A)_{k-1/2}^{n+1/2}(\frac{\partial u}{\partial z})_{k-1/2}^n] \qquad (15)$$

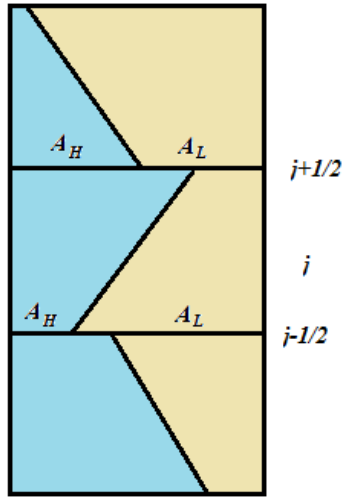

Figure 2 2D schematics showing the surfaces of the heavy and light fluids at the cell faces at *j-1/2* and *j+1/2*.

By considering Figure 2, the above discretisation for stresses can be written as

$$\frac{\Delta t}{V}[(\mu_H A_H + \mu_L A_L)_{i+1/2}^{n+1/2}(\frac{\partial u}{\partial x})_{i+1/2}^n - (\mu_H A_H + \mu_L A_L)_{i-1/2}^{n+1/2}(\frac{\partial u}{\partial x})_{i-1/2}^n$$
$$+ (\mu_H A_H + \mu_L A_L)_{j+1/2}^{n+1/2}(\frac{\partial u}{\partial y})_{j+1/2}^n - (\mu_H A_H + \mu_L A_L)_{j-1/2}^{n+1/2}(\frac{\partial u}{\partial y})_{j-1/2}^n \quad (16)$$
$$+ (\mu_H A_H + \mu_L A_L)_{k+1/2}^{n+1/2}(\frac{\partial u}{\partial z})_{k+1/2}^n - (\mu_H A_H + \mu_L A_L)_{k-1/2}^{n+1/2}(\frac{\partial u}{\partial z})_{k-1/2}^n]$$

In case of large difference in viscosities between the heavy and light fluids, the contribution to the shear stress, say from $(\mu A)_{j+1/2}^{n+1/2}$ in (15), can be more appropriately approximated by $(\mu_H A_H + \mu_L A_L)_{j+1/2}^{n+1/2}$ as that shown in Figure 2. This is different from that given in (1). The equivalent viscosity should thus be that from the surface-area weighted as those given in (16), i.e.

$$\mu = (\mu_H A_H + \mu_L A_L)/A = \mu_H s_H + \mu_L(1 - s_H) \quad (17)$$

rather than that of the volume weighted average as that given in (1). Here *A* is the corresponding surface area of the finite volume. Here $s_H = A_H/A$ is a surface fraction.

In the one-fluid model, the heavy and light fluids in the same cell share the same velocity and the velocity is generally continuous across the interface if no phase change is involved due to the boundary conditions at the interface. Because of this, the $(\mu A)_{j+1/2}^{n+1/2}(\partial u/\partial y)_{j+1/2}^n$ in (15) has been approximated as $(\mu_H A_H + \mu_L A_L)_{j+1/2}^{n+1/2}(\partial u/\partial y)_{j+1/2}^n$ which assumes that $(\partial u/\partial y)_{j+1/2}^n$ is the same at the *j+1/2* surface for both the heavy and light fluids.

The velocity derivative terms at the surfaces between the neighbouring cells in the above discretisation can be approximated by using the velocities at the cell centres, e.g.

$$(\frac{\partial u}{\partial y})_{j+1/2}^n = \frac{u_{j+1}^n - u_j^n}{\Delta y} \quad (18)$$

As shown in Figure 2, the $A_H$ and $A_L$ at *j+1/2* should be the averages between those from cells *(i, j, k)* and *(i, j+1, k)* at the surface *j+1/2* (there could be mismatches at the cell surfaces for the interfaces from the two neighbouring cells since the interface reconstructions in each cell may produce different intersections on each side of the surface at *j+1/2*). This discretization of the viscous terms has automatically taken into account the viscosity jump at the interface, and we believe that this is the appropriate way to account for the viscosity jump at the interface.

The viscous term $\nabla[\mu^{n+1/2}\mathbf{D}^*]$ can be discretised similarly as that for $\nabla[\mu^{n+1/2}\mathbf{D}^n]$ given above. The viscosities of the heavy and light fluids at the surface between the cells are weighted by their corresponding surface areas as in (16) at the intermediate time step $n+1/2$, i.e. $(\mu_H A_H + \mu_L A_L)_{i+1/2}^{n+1/2}$. The discretization of $\mathbf{D}^*$ using $\mathbf{u}^*$ is the familiar Crank-Nicholson scheme and will ensure the scheme in (8) to be second order accurate and unconditionally stable.

As expected, all the spatial discretisation presented here becomes those for single phase flow in case there is no interface intersecting the cell $(i,j,k)$.

## 4. Interface reconstruction

From the above discretisation schemes, it can be seen that the interface between the heavy and light fluids at the cell $(i,j,k)$ needs to be reconstructed to determine its unit normal $\mathbf{n}$, interface area $A_s$ as required in (12), and its intersections with the cell surfaces. These intersections will be used to determine the areas of $A_L$ and $A_H$ as required in (16) as well.

### 4.1 Local second order curve (LSOC) fitting of the interface

The interface reconstruction uses the height function and a second order surface fitting. As an illustration, consider Figure 3 (a 2D case), the volume of fluid can be added column wise to define a height function $h_{ij}= h(x_i, y_j)$. The volume of fluid $c$ used in (1) is actually a volume fraction because it takes values between 0 and 1, and the volume of cell $(i,j,k)$ has been assumed as 1. In numerical implementation, a conversion is needed to transform the volume of fluid $c$ into a local height function $h$.

A locally second order function $z=h(x, y)$ is assumed as
$$h(x,y) = a_0 + a_1 x + a_2 y + a_3 xy + a_4 x^2 + a_5 y^2 \qquad (19)$$
to fit the interface. To determine the coefficients $a_0$ to $a_5$, at least 6 surface cells in the $(x, y)$ plane is required. By considering the surface cells next to $i$ and $j$, i.e. from $i-1$ to $i+1$ and from $j-1$ to $j+1$, we can have up to 9 surface cells. In case that not enough surface cells can be found to determine these coefficients uniquely, i.e. less than 6 surface cells can be found, either the surface fitting needs to change its orientation such as $x=h(y, z)$ (or $y=h(x, z)$) or a local grid refinement is required using an adaptive mesh refinement (AMR) scheme such as the Quad/octree discretisation used in [51]. Also, it is required that the interface intersects with the cell surfaces at both ends. In Figure 3, this means that the interface intersects with the $i+1+1/2$ surface on the right and the $i-1-1/2$ surface on the left. If these attempts failed, a corner (or singularity) has been identified, and the consideration of surface fitting at the corners will be explained later. In the above interface fitting by a curved surface, we have also assumed that the interface is on top of the heavy fluid. If the interface is below the heavy fluid, a negative volume of fluid or the volume of fluid of the light fluid (1-$c$) can be used (in this case the interface is on top of the light fluid) to calculate the height function. Similar consideration applies in case the surface fitting is in the $x$ or $y$ direction.

The function form of (19) is the same as that given in [42] in the parabola-fitted curvature calculation. However, fitting the interface using the height function is not the same as a general curve (or surface) fitting since the height function obtained above is actually an average height within the column $(i, j)$ after summing the volume of fluid in each column, the exact location where the interface achieves this average height is not known, and this cannot be approximated as the centre of the surface cell $(i, j)$. This can be seen from Figure 3, for the top two cells where the volume of fluids are nonzero, the interface may not even exist at the corresponding centres of the surface cells. Thus the quantities that need to be fitted are the average height, i.e. we need a surface (a curved surface with an average unit normal different from that of the $z$ axis) which cuts through the column above the surface cell $(i, j)$ and yields the approximate average height function. This can be achieved by a local integration of the height function (19).

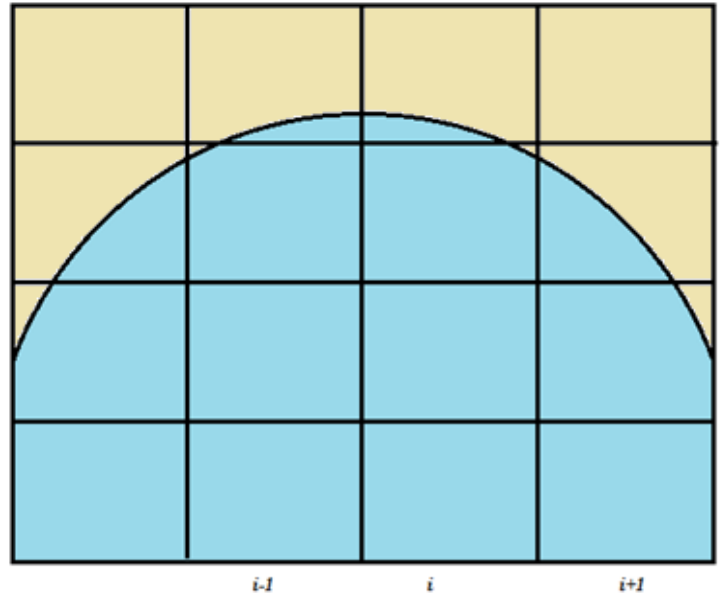

Figure 3 2D schematics to show the calculation of the height function

By integrating (19) locally over the surface cell $(i, j)$, we obtain a volume in the column

$$\int_{y_{j-1/2}}^{y_{j+1/2}} \int_{x_{i-1/2}}^{x_{i+1/2}} h \, dx dy = a_0 \Delta x \Delta y + \frac{1}{2} a_1 \Delta y (x_{i+1/2}^2 - x_{i-1/2}^2) + $$
$$\frac{1}{2} a_2 \Delta x (y_{j+1/2}^2 - y_{j-1/2}^2) + \frac{1}{4} a_3 (x_{i+1/2}^2 - x_{i-1/2}^2)(y_{j+1/2}^2 - y_{j-1/2}^2) + \quad (20)$$
$$\frac{1}{3} a_4 \Delta y (x_{i+1/2}^3 - x_{i-1/2}^3) + \frac{1}{3} a_5 \Delta x (y_{j+1/2}^3 - y_{j-1/2}^3)$$

where $\Delta x = x_{i+1/2} - x_{i-1/2}, \Delta y = y_{j+1/2} - y_{j-1/2}$. Dividing both sides of (20) by $\Delta x \Delta y$, this will result in a locally approximate average height to $h_{ij}$ over the surface cell $(i, j)$ and we thus have

$$\hat{h}_{ij} = a_0 + \frac{1}{2} a_1 (x_{i+1/2} + x_{i-1/2}) + \frac{1}{2} a_2 (y_{j+1/2} + y_{j-1/2}) + $$
$$\frac{1}{4} a_3 (x_{i+1/2} + x_{i-1/2})(y_{j+1/2} + y_{j-1/2}) + \frac{1}{3} a_4 (x_{i+1/2}^2 + x_{i+1/2} x_{i-1/2} + x_{i-1/2}^2) \quad (21)$$
$$+ \frac{1}{3} a_5 (y_{j+1/2}^2 + y_{j+1/2} y_{j-1/2} + y_{j-1/2}^2)$$

If there are exactly six surface cells available for the interface reconstruction, then $\hat{h}_{ij} = h_{ij}$, the coefficients $a_0$ to $a_5$ can be determined uniquely by solving a set of six linear equations. If there are more than six surface cells available, say, $N > 6$ surface cells, a least square curve fitting procedure can be applied. This can be a minimization of the weighted error function

$$E = \sum_{l=j-1}^{j+1} \sum_{m=i-1}^{i+1} w_{ml} (h_{ij} - \hat{h}_{ij})^2 \qquad (22)$$

with $a_0$ to $a_5$ as unknowns. In (22), $w_{ml}$ is a weighting function which depends on the distance from the surface cell centre $(i, j)$. This will be the same as that used in [38] for 3D surface Least-Squares Fit (LSF) method, except here the curve fitting is a height function. Specifically,

$$w'_{ml} = \exp(-\bar{d}^2 / (a\sigma^2)) \qquad (23)$$

$$d_{ml}^2 = (x_{ml} - x_C)^2 + (y_{ml} - y_C)^2 \qquad (24)$$

$$\bar{d} = \sum_{l=j-1}^{j+1} \sum_{m=i-1}^{i+1} d_{ml} / N \qquad (25)$$

$$\sigma^2 = \sum_{l=j-1}^{j+1} \sum_{m=i-1}^{i+1} (d_{ml} - \bar{d})^2 / (N(N-1)) \qquad (26)$$

$$w_T = \sum_{l=j-1}^{j+1} \sum_{i=i-1}^{i+1} w'_{ml} \qquad (27)$$

$$w_{ij} = w'_{ij} / w_T \qquad (28)$$

The constant $a$ in (23) can be taken as 0.75 as that in [38]. With $w_{ij} = 1$, the weightings from all the neighbouring surface cells to $(i, j)$ are the same and it becomes the conventional least square curve fitting.

The function $E$ is minimized by taking

$$\frac{\partial E}{\partial a_0} = \frac{\partial E}{\partial a_1} = \frac{\partial E}{\partial a_2} = \frac{\partial E}{\partial a_3} = \frac{\partial E}{\partial a_4} = \frac{\partial E}{\partial a_5} = 0 \qquad (29)$$

This again will produce six linear equations which can be solved for $a_0$ to $a_5$. In contrast to many of the interface reconstruction methods (including some of the PLIC methods), the LSOC method solves only a set of six linear equations in the cell $(i, j, k)$ where the interface exists, and there is no iterations in reconstructing the interface. So it is expected that the reconstruction of the interface by using LSOC can be fast.

To preserve the volume of fluid at the cell $(i, j, k)$, the value $a_0$ in (21) can be adjusted to achieve $\hat{h}_{ij} = h_{ij}$. If it is found that this adjustment is larger than a certain percentage of $h_{ij}$, a local grid refinement is required. The above fitted interface can be used for all the cells in the column $(i, j)$ in which the interface exists (as that shown in Figure 3, there are two cells above the *i-1*, *i*, and *i+1* columns, respectively, where the interfaces exist). This means that only one interface needs to be reconstructed for the two cells in each of the columns, so there is a possible time saving in implementing the LSOC mrthod. In case that the interface passes through three or more cells above each column, a change of orientation in fitting the interface needs to be considered or the cell can be considered as a corner (to be discussed later).

The area of the interface in the cell (*i, j, k*) is calculated as

$$A_s = \int_{y_{j-1/2}}^{y_{j+1/2}} \int_{x_{i-1/2}}^{x_{i+1/2}} g(x,y) dx dy \tag{30}$$

where

$$g(x,y) = \begin{cases} \sqrt{1+z_x^2+z_y^2} & \text{if } z_{k-1/2} < h < z_{k+1/2} \\ 0 & \text{otherwise} \end{cases}$$

$$h_x = \frac{\partial h}{\partial x} = a_1 + a_3 y + 2a_4 x \tag{31}$$

$$h_y = \frac{\partial h}{\partial y} = a_2 + a_3 x + 2a_5 y$$

Locally this surface integration can be calculated, say using *dx*=Δ*x*/10 and *dy*=Δ*y*/10, so a more accurate surface area can be calculated. The curvature on the interface can be calculated [52] as

$$H = \frac{1}{2}(\kappa_1 + \kappa_2) = \frac{(1+h_x^2)h_{yy} - 2h_x h_y h_{xy} + (1+h_y^2)h_{xx}}{(1+h_x^2+h_y^2)^{3/2}} \tag{32}$$

where $h_{yy} = 2a_5$, $h_{xx} = 2a_4$ and $h_{xy} = a_3$. For a linear plane surface, *H* = 0.

The unit normal on the interface can be calculated [52] as

$$\mathbf{n} = (\frac{h_x}{\sqrt{1+h_x^2+h_y^2}}, \frac{h_y}{\sqrt{1+h_x^2+h_y^2}}, \frac{-1}{\sqrt{1+h_x^2+h_y^2}}) \tag{33}$$

### 4.2  Determining the forces due to the surface tension

The force from the surface tension at the interface can be calculated from the unit normal (33) and the curvature (32) as shown in (12).  However, the unit normal and the curvature of the interface vary within the cell $(i,j,k)$ for a second order curved surface. Because of these, the average unit normal and the curvature within the cell $(i,j,k)$ need to be calculated and these cannot in general be taken as the values of (33) and (32) at the centre of $(i,j)$ since the interface can intersect with the cells in a complex manner as those shown later in Figure 6. One way to calculate the forces due to the surface tension on the interface is to take a local surface integration and (12) can be calculated as

$$\frac{\Delta t}{V}(\int_{A_s} \sigma \kappa \mathbf{n} dS)_{i,j,k} = \frac{2\sigma \Delta t}{V} \int_{y_{j-1/2}}^{y_{j+1/2}} \int_{x_{i-1/2}}^{x_{i+1/2}} H\mathbf{n} g(x,y) dx dy \tag{34}$$

where the *g(x, y)*, *H*, and **n** are given in (31), (32) and (33), respectively. Here we have assumed that the surface tension $\sigma$ is constant within the cell $(i,j,k)$.

The force from the surface tension at the interface within the cell $(i,j,k)$ can also be calculated from the line integration along the intersection lines from the definition of the surface tension [53]. Figure 4 shows a schematic diagram of the interface S with a bounding contour C on the interface between the two fluids with local unit vectors, **n**, **m** and **s**.

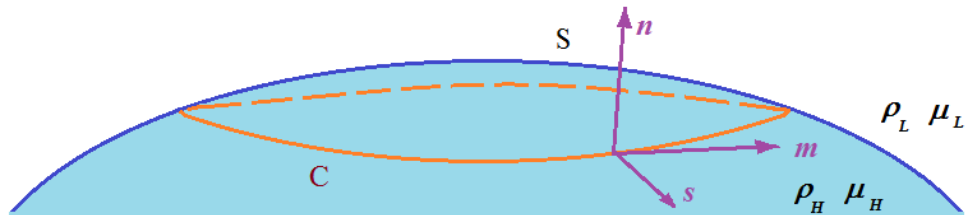

Figure 4 A surface S and bounding contour C on an interface between the heavy and light fluids. Local unit vectors are **n**, **m** and **s**.

For the bounding contour C shown in Figure 4, the force due to the surface tension can be calculated as [53]

$$\oint_C \sigma \mathbf{s} dl \tag{35}$$

Figure 5 shows some intersections of the interface with the surfaces of the cell $(i,j,k)$ when the interface is close to a plane surface as those illustrated in [38]. For simplicity, only straight intersection lines are drawn on the cell faces. For curved interfaces, more complex intersections with the cell surfaces can be formed as that shown in Figure 6, which shows that locally fitting these interfaces with a plane surface would result in large errors.

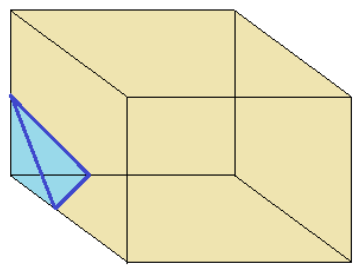

a) 3 intersections

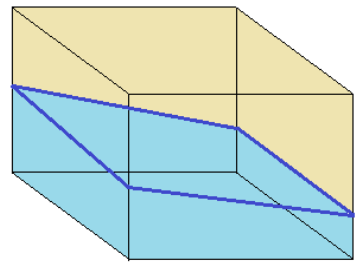

b) 4 intersections

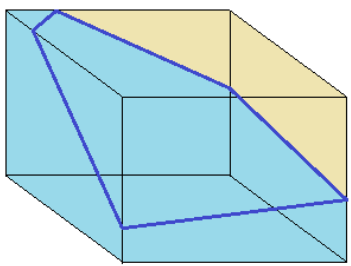

c) 5 intersections

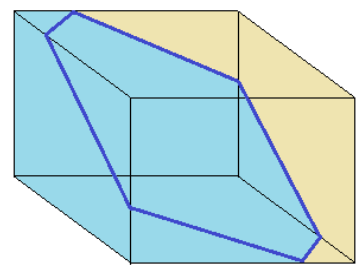

d) 6 intersections

Figure 5 Possible intersection curves with the cell centred at (*i, j, k*) by a plane. The number of intersection curves varies from three to six (from [38]).

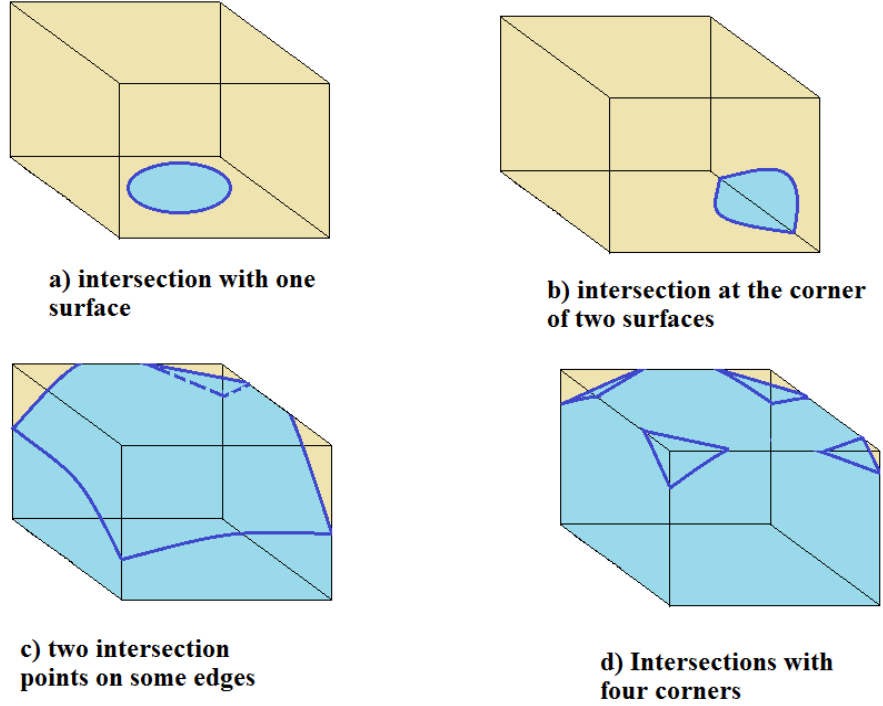

a) intersection with one surface

b) intersection at the corner of two surfaces

c) two intersection points on some edges

d) Intersections with four corners

Figure 6 Possible intersection curves with the cell centred at (*I, j, k*) by a curved surface

For the intersection curves of the interface with the cell surfaces as those shown in Figures 5 and 6, the vector **m** lies in one of the cell surfaces. Its direction should be chosen in such a way that

$$\mathbf{s} = \mathbf{m} \times \mathbf{n} \tag{36}$$

and **s** is pointing away from the interface enclosed by the contour C and towards the outside of the cell $(i, j, k)$. The circle integration in (35) becomes the integrations along each of the intersection curves as shown in Figures 5 and 6. As shown in Figure 6, some of the intersection curves can be quite complex.

By applying Stokes' Theorem, the line integration in (35) can be converted into the surface integration (34).

### 4.3 Surface fitting near wall boundaries

At wall boundary conditions, the surface fitting at the cells next to the walls can be performed in one of the following two methods.

One method is that, for the cells near the wall, if six cells can be found, then the coefficients in (19) can be determined uniquely. In Figure 3, assuming the cells with *i*+1 are next to a wall (with the wall at its right surface), then the six cells can be (*i, j*-1), (*i, j*), (*i, j*+1), (*i*+1, *j*-1), (*i*+1, *j*), (*i*+1, *j*+1). If less than six cells have been found, a local grid refinement is needed.

In case that honouring the static contact angle at the walls is important, the second method can be used to determine the surface fitting. In this case, the interface normal at the wall and the wall surface normal can be assumed to satisfy the static contact angle $\theta_0$ [40], i.e.

$$\hat{\mathbf{n}}_f \cdot \hat{\mathbf{n}}_w = \cos\theta_0 \tag{37}$$

where $\hat{\mathbf{n}}_f$ and $\hat{\mathbf{n}}_w$ are the unit normal of the interface and that of the wall surface at the intersection of the wall surface and the interface. Using (33) and (35), this results in a constrain on the height function at, say $x_{i+1/2}$, assuming the cells with $i+1$ are next to a wall. This will result in a nonlinear surface fitting process which we will not discuss here.

### 4.4 Surface fitting near corners

In multi-phase flows, singularities can exist such as water droplets break away from a liquid jet. At the time the singularity is formed, the interface at the singularity between the heavy and light fluids will not be smooth and this can be considered as corners. In general, the corners cannot be resolved by grid refinement and need to be treated specially. It is also impossible to capture the formation of this singularity in time and space location numerically. Figure 7 shows an interface with a corner in cell $(i+1, j+1)$. In Figure 7, the interface in cells $(i-1, j+1)$ and $(i+1, j-1)$ can be determined using the procedures presented early (with different orientations of the height functions) using the LSOC method. The interfaces in the cells $(i, j+1)$ and $(i+1, j)$ can be determined using the six cells available as that of method 1 presented in **4.3** for fitting interfaces near wall boundaries. The interface in cell $(i+1, j+1)$ can then be approximated by extrapolating the interfaces from $(i, j+1)$ and $(i+1, j)$ cells and the corner location can be determined from these extrapolations. It is expected that, at the time the singularity occurs, some discontinuities in pressure and velocity (more specifically, $\partial p/\partial t$ and $\partial \mathbf{u}/\partial t$) could be generated. Since the time and spatial locations of the singularities cannot be captured in numerical simulations, there is currently no satisfactory theory to account for these effects.

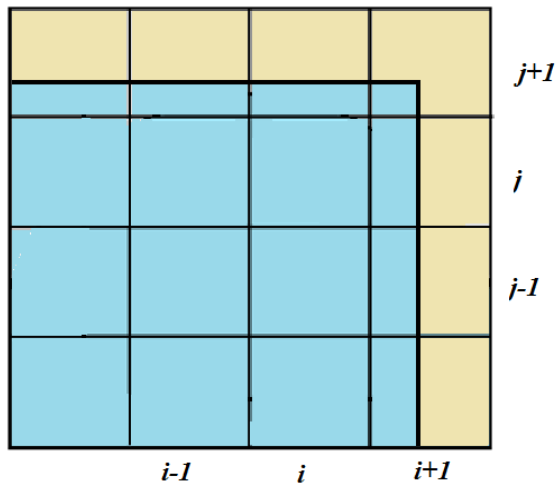

Figure 7 interface with corners (singularities)

## 5. Determining the mass advections $m_H$ and $m_L$

The discretisation in (11) and (14) requires the calculation of advection masses of the heavy and light fluids from the upstream cells. Here we will adapt a similar strategy as that in [34] to determine these masses using the fitted second order surface. For simplicity, we will use the directional split scheme for the mass advections. As in [34], for a given cell $(i,j,k)$, we will determine its departure region $\Omega_{i,depart}$ as that shown in Figure 8 in the $x$-direction with the target region

$$\Omega_i = \{x_{i-1/2}, x_{i+1/2}\} \tag{38}$$

The departure region can be approximated as

$$\Omega_{i,depart} = \{x_{i-1/2} - \Delta\tilde{x}_{i-1/2}, x_{i+1/2} - \Delta\tilde{x}_{i+1/2}\} \tag{39}$$

By assuming a linear variation of the velocity between the neighbouring cell centres, the $\Delta\tilde{x}_{i-1/2}$ and $\Delta\tilde{x}_{i+1/2}$ in (39) can be determined (assuming the grid size in the $x$ direction is uniform) as

$$\Delta\tilde{x}_{i-1/2} = \frac{0.5(u_{i-1}+u_i)\Delta t}{1+\dfrac{u_i-u_{i-1}}{x_i-x_{i-1}}\Delta t} \approx \frac{0.5(u_{i-1}+u_i)\Delta t}{1+\dfrac{\partial u}{\partial x}\big|_{i-1/2}\Delta t} \tag{40}$$

$$\Delta\tilde{x}_{i+1/2} = \frac{0.5(u_i+u_{i+1})\Delta t}{1+\dfrac{u_{i+1}-u_i}{x_{i+1}-x_i}\Delta t} \approx \frac{0.5(u_i+u_{i+1})\Delta t}{1+\dfrac{\partial u}{\partial x}\big|_{i+1/2}\Delta t} \tag{41}$$

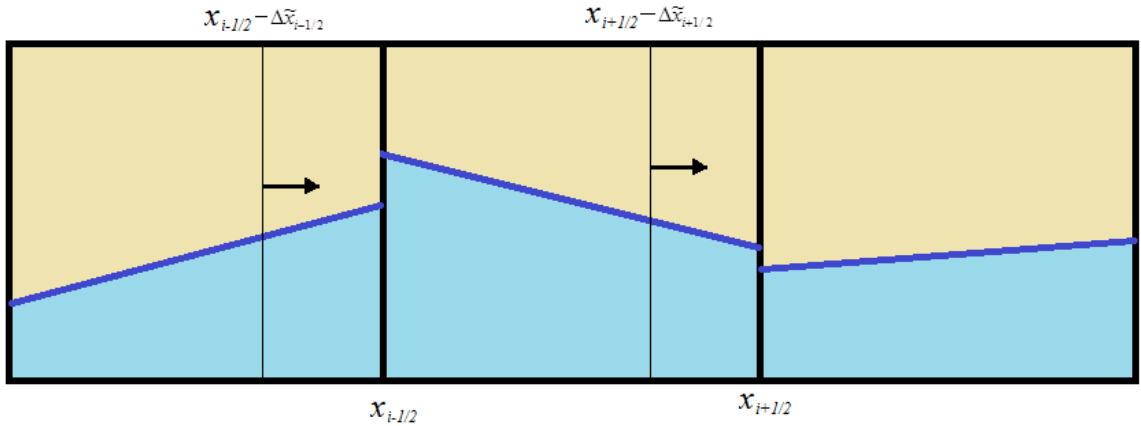

Figure 8 The departure region and the target region.

The mass of the heavy fluid advected through the cell surface between $(i-1,j,k)$ and $(i,j,k)$ can be calculated as

$$(m_H)_{i-1/2} = \rho_H(1+\frac{\partial u}{\partial x}\big|_{i-1/2}\Delta t)\int_{x_{i-1/2}-\Delta\tilde{x}_{i-1/2}}^{x_{i-1/2}}\int_{y_{j-1/2}}^{y_{j+1/2}} f_{i-1}dxdy \tag{42}$$

$$f_{i-1} = \begin{cases} h_{i-1} - z_{k-1/2} & z_{k-1/2} < h_{i-1} < z_{k+1/2} \\ 0 & h_{i-1} < z_{k-1/2} \\ z_{k+1/2} - z_{k-1/2} & h_{i-1} > z_{k+1/2} \end{cases} \tag{43}$$

where $h_{i-1}$ is the fitted surface (19) for the interface in the cell $(i-1, j, k)$. The factor $(1+\frac{\partial u}{\partial x}|_{i-1/2} \Delta t)$ in (42) is to take into account the stretching ( $(\partial u / \partial x)|_{i-1/2} > 0$ ) or compression ( $(\partial u / \partial x)|_{i-1/2} < 0$ ) of the fluid as it is advected from the cell $(i-1, j, k)$ to cell $(i, j, k)$.

Similarly, the mass of the light fluid advected through the cell surface between $(i-1, j, k)$ and $(i, j, k)$ can be calculated as

$$(m_L)_{i-1/2} = \rho_L (1+\frac{\partial u}{\partial x}|_{i-1/2} \Delta t) \int_{x_{i-1/2}-\Delta \tilde{x}_{i-1/2}}^{x_{i-1/2}} \int_{y_{j-1/2}}^{y_{j+1/2}} g_{i-1} dx dy \tag{44}$$

$$g_{i-1} = \begin{cases} z_{k+1/2} - h_{i-1} & z_{k-1/2} < h_{i-1} < z_{k+1/2} \\ z_{k+1/2} - z_{k-1/2} & h_{i-1} < z_{k-1/2} \\ 0 & h_{i-1} > z_{k+1/2} \end{cases} \tag{45}$$

In case that $\Delta \tilde{x}_{i-1/2} = 0$, which means $0.5(u_i + u_{i-1}) = 0$, the above expressions show that $(m_H)_{i-1/2} = 0$ and $(m_L)_{i-1/2} = 0$, as expected.

The calculations of the mass advections in the $y$ direction are similar to those of the $x$ direction and will not be repeated here. For the mass advections in the $z$ direction,

$$(m_H)_{k-1/2} = \rho_H (1+\frac{\partial w}{\partial z}|_{k-1/2} \Delta t) \int_{x_{i-1/2}}^{x_{i+1/2}} \int_{y_{j-1/2}}^{y_{j+1/2}} f_{k-1} dx dy \tag{46}$$

with

$$f_{k-1} = \begin{cases} 0 & h_{k-1} < z_{k-1/2} - \Delta \tilde{z}_{k-1/2} \\ h_{k-1} - (z_{k-1/2} - \Delta \tilde{z}_{k-1/2}) & z_{k-1/2} - \Delta \tilde{z}_{k-1/2} < h_{k-1} < z_{k-1/2} \\ \Delta \tilde{z}_{k-1/2} & h_{k-1} > z_{k-1/2} \end{cases} \tag{47}$$

where $h_{k-1}$ is the fitted surface for the interface in the cell $(i, j, k-1)$ and

$$\Delta \tilde{z}_{k-1/2} = \frac{0.5(w_{k-1}+w_k)\Delta t}{1+\frac{w_k - w_{k-1}}{z_k - z_{k-1}}\Delta t} \approx \frac{0.5(w_{k-1}+w_k)\Delta t}{1+\frac{\partial w}{\partial z}|_{k-1/2} \Delta t} \tag{48}$$

for uniform grid size in the $z$ direction. The corresponding $(m_L)_{k-1/2}$ can be calculated as

$$(m_L)_{k-1/2} = \rho_L (1+\frac{\partial w}{\partial z}|_{k-1/2} \Delta t) \int_{x_{i-1/2}}^{x_{i+1/2}} \int_{y_{j-1/2}}^{y_{j+1/2}} g_{k-1} dx dy \tag{49}$$

$$g_{k-1} = \begin{cases} \Delta \tilde{z}_{k-1/2} & h_{k-1} < z_{k-1/2} - \Delta \tilde{z}_{k-1/2} \\ z_{k-1/2} - h_{k-1} & z_{k-1/2} - \Delta \tilde{z}_{k-1/2} < h_{k-1} < z_{k-1/2} \\ 0 & h_{k-1} > z_{k-1/2} \end{cases} \tag{50}$$

The mass advections $m_H$ and $m_L$ in the mass conservation (11) can be kept for calculating the momentum advection in (14), and thus they only need to be calculated once during each iteration loop.

As mentioned in *Remark 2*, the velocity $u_{i-1/2}^{n+1/2}$, etc, can be approximated using various methods. The simplest method is

$$\widetilde{u}_{i-1/2}^{n+1/2} \approx 0.5(u_{i-1}^{n+1/2} + u_i^{n+1/2}) \tag{51}$$

Here we have assumed that the grid size in the *x* direction is uniform and the velocity between neighbouring cells varies linearly. This approximation of the surface velocity is similar to that of the second order upwind.

The second method of approximating this velocity is

$$\widetilde{u}_{i-1/2}^{n+1/2} \approx u(x_{i-1/2} - 0.5 u_{i-1/2}^{n+1/2} \Delta t) \tag{52}$$

Here the $u_{i-1/2}^{n+1/2}$ is that given in (51).

The third method of the approximation is given by

$$\widetilde{u}_{i-1/2}^{n+1/2} \approx u(x_{i-1/2} - 0.5 \Delta \widetilde{x}_{i-1/2}^{n+1/2}) \tag{53}$$

where $\Delta \widetilde{x}_{i-1/2}$ is that given by (40). It should be noted that the methods (2) and (3) of approximating $u_{i-1/2}^{n+1/2}$ can also be used for single phase flow simulations to achieve a higher order accuracy in the discretisation of the momentum equations.

Equations (40) and (42) show that for single phase flows, say for the heavy fluid with a constant density, the mass entering the cell $(i,j,k)$ through the $A_{i-1/2}$ surface is $(m_H)_{i-1/2}^{n+1/2} = \rho_H u_{i-1/2}^{n+1/2} \Delta t A_{i-1/2}$ and the momentum entering the cell $(i,j,k)$ is $\rho_H u_{i-1/2}^{n+1/2} \widetilde{u}_{i-1/2}^{n+1/2} \Delta t A_{i-1/2}$. Here $u_{i-1/2}^{n+1/2} = 0.5(u_{i-1}^{n+1/2} + u_i^{n+1/2})$ for uniform grid and $\widetilde{u}_{i-1/2}^{n+1/2}$ can be one of the approximations from (51) to (53).

For two phase flows, the forth method of approximating the velocity $u_{i-1/2}^{n+1/2}$ can be expressed as

$$(\widetilde{u}_{i-1/2}^{n+1/2})_H \approx u(x_{i-1/2,H}^{n+1/2}) \tag{54}$$

$$(\widetilde{u}_{i-1/2}^{n+1/2})_L \approx u(x_{i-1/2,L}^{n+1/2}) \tag{55}$$

where

$$x_{i-1/2,H}^{n+1/2} = \frac{\int_{x_{i-1/2}-\Delta \widetilde{x}_{i-1/2}}^{x_{i-1/2}} \int_{y_{j-1/2}}^{y_{j+1/2}} x f_{i-1} dx dy}{\int_{x_{i-1/2}-\Delta \widetilde{x}_{i-1/2}}^{x_{i-1/2}} \int_{y_{j-1/2}}^{y_{j+1/2}} f_{i-1} dx dy} \tag{56}$$

$$x_{i-1/2,L}^{n+1/2} = \frac{\int_{x_{i-1/2}-\Delta \widetilde{x}_{i-1/2}}^{x_{i-1/2}} \int_{y_{j-1/2}}^{y_{j+1/2}} x g_{i-1} dx dy}{\int_{x_{i-1/2}-\Delta \widetilde{x}_{i-1/2}}^{x_{i-1/2}} \int_{y_{j-1/2}}^{y_{j+1/2}} g_{i-1} dx dy} \tag{57}$$

with $f_{i-1}$ and $g_{i-1}$ as given in (47) and (50). For single phase flows, only one of the approximations (54) and (55) is required and it is the same as that using (53) for fluids with constant density. With the approximate velocities expressed in (54) and (55), the momentum entering the cell $(i, j, k)$ from the upstream surface $(i-1/2, j, k)$ can be expressed as

$$(m_L + m_H)_{i-1/2}^{n+1/2} u_{i-1/2}^{n+1/2} \approx (m_H)_{i-1/2}^{n+1/2} u(x_{i-1/2,H}^{n+1/2}) + (m_L)_{i-1/2}^{n+1/2} u(x_{i-1/2,L}^{n+1/2}) \tag{58}$$

This last approximation method for $u_{i-1/2}^{n+1/2}$ in (54) and (55) is called the moments of fluid and $x_H$ and $x_L$ are the centres of moments of the heavy and light fluids [34].

For single phase flows with variable density, the momentum entering a cell can also be approximated by (58) with only one mass advection and one approximate velocity at the cell surface. The density variation between neighbouring cells will cause both the mass advection and the centre of moment favouring the region of higher density.

### 6. Determining the surface areas $A_H$ and $A_L$

As shown in Figures 5 and 6, the interface can intersect with the cell surfaces in some complex manner and here we present methods of calculating the $A_H$ and $A_L$ as required in (16) to correctly account for the stresses due to the viscosities by using the fitted surface (19). On each cell surface, only $A_H$ (or $A_L$) needs to be calculated, the $A_L$ can be calculated from $A - A_H$ where $A$ is the surface area at the corresponding cell surface. Here we present the methods of calculating the $A_H$ only. These surfaces need only to be calculated once at each iteration loop, after the final interface reconstruction, irrespective a directional split scheme or an unsplit scheme is used for the advection of the interface.

On the surface $x_{i-1/2}$,

$$A_{H,i-1/2} = \int_{y_{j-1/2}}^{y_{j+1/2}} f_{i-1/2} dy \tag{59}$$

where

$$f_{i-1/2} = \begin{cases} h(x_{i-1/2}, y) - z_{k-1/2} & z_{k-1/2} < h < z_{k+1/2} \\ 0 & h < z_{k-1/2} \\ z_{k+1/2} - z_{k-1/2} & h > z_{k+1/2} \end{cases} \tag{60}$$

and $h(x_{i-1/2}, y)$ is the fitted height function (19) at $x = x_{i-1/2}$.

On the surface $x_{i+1/2}$,

$$A_{H,i+1/2} = \int_{y_{j-1/2}}^{y_{j+1/2}} f_{i+1/2} dy \tag{61}$$

where

$$f_{i+1/2} = \begin{cases} h(x_{i+1/2}, y) - z_{k-1/2} & z_{k-1/2} < h < z_{k+1/2} \\ 0 & h < z_{k-1/2} \\ z_{k+1/2} - z_{k-1/2} & h > z_{k+1/2} \end{cases} \tag{62}$$

and $h(x_{i+1/2}, y)$ is the fitted height function (19) at $x = x_{i+1/2}$.

On the surface $y_{j-1/2}$,

$$A_{H,j-1/2} = \int_{x_{i-1/2}}^{x_{i+1/2}} f_{j-1/2} dx \qquad (63)$$

where

$$f_{j-1/2} = \begin{cases} h(x, y_{j-1/2}) - z_{k-1/2} & z_{k-1/2} < h < z_{k+1/2} \\ 0 & h < z_{k-1/2} \\ z_{k+1/2} - z_{k-1/2} & h > z_{k+1/2} \end{cases} \qquad (64)$$

and $h(x, y_{j-1/2})$ is the fitted height function (19) at $y = y_{j-1/2}$.

On the surface $y_{j+1/2}$,

$$A_{H,j+1/2} = \int_{x_{i-1/2}}^{x_{i+1/2}} f_{j+1/2} dx \qquad (65)$$

where

$$f_{j+1/2} = \begin{cases} h(x, y_{j+1/2}) - z_{k-1/2} & z_{k-1/2} < h < z_{k+1/2} \\ 0 & h < z_{k-1/2} \\ z_{k+1/2} - z_{k-1/2} & h > z_{k+1/2} \end{cases} \qquad (66)$$

and $h(x, y_{j+1/2})$ is the fitted height function (19) at $y = y_{j+1/2}$.

On the surface $z_{k-1/2}$,

$$A_{H,k-1/2} = \int_{x_{i-1/2}}^{x_{i+1/2}} \int_{y_{j-1/2}}^{y_{j+1/2}} f_{k-1/2} dx dy \qquad (67)$$

where

$$f_{k-1/2} = \begin{cases} 1 & h > z_{k-1/2} \\ 0 & h < z_{k-1/2} \end{cases} \qquad (68)$$

On the surface $z_{k+1/2}$,

$$A_{H,k+1/2} = \int_{x_{i-1/2}}^{x_{i+1/2}} \int_{y_{j-1/2}}^{y_{j+1/2}} f_{k+1/2} dx dy \qquad (69)$$

where

$$f_{k+1/2} = \begin{cases} 1 & h > z_{k+1/2} \\ 0 & h < z_{k+1/2} \end{cases} \qquad (70)$$

Again, the local integration of these surface areas can be carried out using, say, $dx=\Delta x/10$ and $dy=\Delta y/10$.

## 7. Discussion and Conclusions

A comprehensive numerical scheme at the interface of the two fluids based on the finite volume and VOF methods is presented for numerically simulating incompressible two phase flows with large density and viscosity differences. The numerical scheme is based on the physical requirements of mass conservation, momentum balance and proper account of the shear and normal stresses contributed by the heavy and light fluids in a finite volume. The interface advection is determined by the continuity equation to ensure that the mass conservation is satisfied. Since in a flow field, any mass flow carries with it the momentum, the advection scheme for the heavy and light fluids at the interfaces are used to determine the advection of the momentum in order to achieve a consistent numerical scheme to discretise both the continuity and momentum equations. By considering the force balances on the surfaces of a finite volume, it is shown that the normal and shear stresses on the surfaces need to be calculated using a surface-area weighted viscosity rather than a volume weight viscosity.

We have presented a locally second order surface fitting method LSOC for the interface reconstruction based on the height functions. The scheme takes into account the fact that the height function calculated by adding the volume of fluid in a column produces only the average height in the column, with the detail information on the variation of the interface inside the cell being lost. In order to curve fitting this average height in each cell, the continuous height function to be determined is locally integrated first so an average height can be determined. Thus the LSOC interface reconstruction scheme is mathematically consistent. The interface reconstruction scheme is linear and in general only six linear equations need to be solved in each reconstruction.

With the fitted surface for the interface, the curvature, the surface area and the unit normal of the interface can all be determined analytically and thus the forces due to the surface tension can be calculated more accurately since no discretisation of a step function is involved.

We have also presented the methods of calculating the surface areas $A_H$ and $A_L$ on the surfaces of the finite volume using the reconstructed interface function, and the methods of calculating the mass advections $m_H$ and $m_L$.

Several methods of various accuracies are proposed to approximate the velocity for the momentum advection across the surfaces of the finite volume. It is believed that the consideration of the departure region and target region can help to develop even higher order discretisation schemes for the momentum equations in case higher order velocity variation between cell centres is assumed. This idea applies not only for the simulation of two phase flows, but it can also be applied to the simulations of single phase flows.

In the schemes present here, the calculation of the forces due to the surface tension, the local surface area $A_H$ and $A_L$ and the masses of advection $m_H$ and $m_L$ are all based on local integrations. These integrations can improve the accuracy of the schemes but may require more simulation time. This needs to be checked and compared with the conventional schemes available.

The proposed LSOC scheme assumes that the numerical resolution of the interface is sufficient and needs to be combined with an adaptive mesh refined (AMR) scheme such as that used in [51]. Near singularities, the interface has been reconstructed by the extrapolations of reconstructed interfaces from the neighbouring cells.

These schemes need to be verified by numerically simulating two phase flows and compared with experimental data if possible.